\documentstyle[aps,prd,amssymb]{revtex}
\input epsfig.sty
\def\laq{\raise 0.4ex\hbox{$<$}\kern -0.8em\lower 0.62 ex\hbox{$\sim$}}
\def\gaq{\raise 0.4ex\hbox{$>$}\kern -0.7em\lower 0.62 ex\hbox{$\sim$}}
\def\dis{\displaystyle}
\def\bdis{\begin{displaymath}}
\def\edis{\end{displaymath}}
\draft

\begin{document}

\title{Sensitivity of wide band detectors to quintessential gravitons}

\author{ D. Babusci$^{(a)}$ 
\footnote{Electronic address:Danilo.Babusci@lnf.infn.it }  
and M. Giovannini$^{(b)}$\footnote{
Electronic address: giovan@cosmos2.phy.tufts.edu} }

\address{{\it $^{(a)}$ INFN - Laboratori Nazionali di 
Frascati, 00044 Frascati, Italy}}

\address{{\it $^{(b)}$ Department of Physics 
and Astronomy, Tufts University, Medford, Massachusetts 02155, USA}}

\maketitle
\begin{abstract}
There are no reasons why the energy spectra 
of the relic gravitons amplified by the pumping action of the background 
geometry should not increase at high frequencies. A
typical example of this behavior are  quintessential inflationary
models where the slopes of the energy spectra can be either blue 
or mildly violet. In comparing the predictions of scenarios leading to 
blue and violet graviton spectra we face the problem of correctly 
deriving the sensitivities of the interferometric detectors.
Indeed the expression of the signal-to-noise ratio not only 
depends upon the noise power spectra of the detectors but 
also upon the spectral form of the signal and, therefore, one can 
reasonably expect that models with different spectral 
behaviors will produce different signal-to-noise ratios. 
By assuming monotonic (blue) spectra of relic gravitons we 
will give general expressions for the  signal-to-noise ratio
in this class of models. As an example we studied the case 
of quintessential gravitons. The minimum achievable 
sensitivity to $h^2_{0}\,\Omega_{{\rm GW}}$ of different pairs 
of detectors is computed, and compared with the theoretical 
expectations.
\end{abstract}

\renewcommand{\theequation}{1.\arabic{equation}}
\setcounter{equation}{0}
\section{Introduction}

Gravitational wave astronomy, experimental cosmology and 
high energy physics will soon experience 
a boost thanks to the forthcoming interferometric detectors. 
From a theoretical  point of view it is then 
interesting to compare our theoretical expectations/speculations 
with the foreseen sensitivities of the various devices in 
a frequency range 
which complements and greatly extends the information we can 
derive from the analysis of the microwave sky and of its 
temperature fluctuations. 

By focusing our attention on relic gravitons of primordial origin 
we can say that 
virtually every variation in the time evolution of the 
curvature scale can imprint important informations on the stochastic 
gravitational wave background \cite{1}. The problem is that the precise 
evolution of the curvature scale is not known. Different 
cosmological scenarios, based on different physical models of the early 
Universe may lead to different energy spectra of relic gravitons 
and this crucial theoretical indetermination can 
affect the expected signal. 

Of particular interest seems to be the case where the logarithmic energy 
density of the relic gravitons (in critical units) grows \cite{2,3}
in the frequency region explored by the interferometric 
detectors (i.e., approximately between few Hz and $10$ kHz) 
\cite{4,5,6,7,8}.
In this range we can parametrize the energy density of the 
relic gravitons $\rho_{{\rm GW}}$ at the present time $\eta_0$ as 
\begin{equation}
\Omega_{{\rm GW}}(f,\eta_0) = \frac{1}{\rho_{c}} \frac{d \rho_{{\rm GW}}}{d 
\ln{f}}= \overline{\Omega}(\eta_0) q(f,\eta_0) 
\label{1}
\end{equation}
where $\overline{\Omega}(\eta_0)$ denotes the typical amplitude of the 
spectrum and $q(f,\eta_0)$ is a monotonic function of the frequency 
at least in the interval $1 ~{\rm Hz}\;\laq\;f\;\laq\;10 ~{\rm kHz}$. Both 
$\overline{\Omega}(\eta_0)$ and 
$q(f,\eta_0)$ can depend on the parameters of the particular 
model. The assumption that  $q(f,\eta_0)$ is  monotonic can certainly be seen
 as  a restriction of our analysis, but, at the same time, we can notice that
the models with growing logarithmic energy spectra which were discussed up 
to now in the litterature fit in our choice for $q(f,\eta_0)$.
Within the parametrization defined in Eq. (\ref{1}) we will be discussing 
the cases where  the spectral slope $\alpha$ 
(i.e., $\alpha = d q(f,\eta_0)/d f$) is
 either blue (i.e., $0< \alpha\;\laq\;1$) or violet (i.e., $\alpha > 1$). 
In general we could have also the case $\alpha < 0$ (red spectra) and 
$\alpha =0$ (flat spectrum). Flat spectra have been 
extensively studied in the context of  ordinary inflationary 
models \cite{9} and in relation to cosmic string models \cite{10}. 

Blue and violet spectra are physically peculiar since they are 
typically produced in models which are 
different from the ones leading to flat spectra.  
In quintessential inflationary models \cite{11} 
the logarithmic energy spectra are 
typically blue \cite{12}. 
This is due to the fact that in this class of models an 
ordinary inflationary phase is  followed by an expanding phase 
whose dynamics is driven  by an effective equation of state which is stiffer
 than radiation \cite{13}. 
Since the equation of state (after the end of inflation) is stiffer 
than the one of radiation, then the Universe will expand slower than in a 
radiation dominated phase and, therefore, $\alpha$ turns out to be at 
most one (up to logarithmic corrections). 

In string cosmological
models \cite{14}
 the graviton spectra can be either blue (if the physical scale 
corresponding to a present frequency of $100$ Hz went out of the horizon
during the string phase) or violet (if the relevant scale crossed the 
horizon during the dilaton driven phase). 

The purpose of this paper is to analyze the sensitivity 
of pairs of interferometric detectors to blue and violet spectra of relic 
quintessential gravitons. The reason 
for such an exercise is twofold. On one hand violet and blue spectra, owing 
to their growth in frequency, might
provide signals which are larger than in the case 
of flat inflationary spectra. On the other hand the 
sensitivity to blue spectra from quintessential inflation 
can be different from the 
one computed in the case of flat spectra from ordinary inflationary models. 
Indeed, 
it is sometimes common practice to compare the theoretical energy density
of the produced gravitons with the sensitivity of various interferometers 
to a flat spectrum. This is, strictly speaking, arbitrary even if, 
sometimes this procedure might lead to correct order of magnitude estimates.

In order to illustrate qualitatively this point let us consider the general
expression of the signal-to-noise ratio (SNR) in the case of correlation 
of two detectors of arbitrary geometry for an observation time $T$. By 
assuming that the intrinsic 
 noises of the detectors are stationary, gaussian, 
uncorrelated, much larger in amplitude than the gravitational strain, and 
statistically independent on the strain itself, one has 
\cite{17,18,19,40}:\footnote{Notice that, with this definition, 
the SNR turns out to be the square 
root of the one used
 in refs. \cite{17,18,19,40}. The reason for our definition lies in the 
remark that the cross-correlation between the outputs $s_{1,2} (t)$ of the 
detectors is defined as:
\bdis
S\,=\,\int_{-T/2}^{T/2}\,{\rm d} t\,\int_{-T/2}^{T/2}\,{\rm d} t'\,
s_1 (t)\,s_2 (t')\,Q (t,t'), 
\edis
where $Q$ is a filter function. Since $S$ is quadratic in the signals, with 
the usual definitions, it contributes to the SNR squared.}
\begin{equation}
{\rm SNR}^2 \,=\,\frac{3 H_0^2}{2 \sqrt{2}\,\pi^2}\,F\,\sqrt{T}\,
\left\{\,\int_0^{\infty}\,{\rm d} f\,\frac{\gamma^2 (f)\,\Omega_{{\rm GW}}(f)}
{f^6\,S_n^{\,(1)} (f)\,S_n^{\,(2)} (f)}\,\right\}^{1/2}\; ,
\label{2}
\end{equation}
($H_0$ is the present value of the Hubble parameter and $F$ depends upon 
the geometry of the two detectors; in the case of the correlation between 
two interferometers $F=2/5$). 
In Eq. (\ref{2}), $S_n^{\,(k)} (f)$ is the (one-sided) noise power 
spectrum of the $k$-th 
$(k = 1,2)$ detector, while $\gamma(f)$ is the overlap reduction function 
\cite{19,40} which is determined by the relative locations and orientations 
of the two
detectors. This function cuts off (effectively) the integrand at a frequency 
$f \sim 1/2d$, where $d$ is the separation between the two detectors.

From eq. (\ref{2}) we can see that the frequency dependence 
of the signal directly enters in the determination of the SNR and, therefore,
we can expect different values of the integral depending upon the relative 
frequency dependence of the signal and of the noise  power spectra 
associated with the detectors. Hence, in order to get precise 
information on the sensitivities of various detectors to blue 
and violet spectra we have to evaluate the SNR for each specific 
model at hand.

The analysis of the SNR is certainly compelling if we want to confront 
quantitatively our theoretical conclusions with the forthcoming 
data. Owing to the differrence among the various 
logarithmic energy spectra of the relic gravitons we can wonder
if different detector pairs can be more or less sensitive 
to a specific theoretical model. We will try, when possible, to state 
our conclusions in such a way that our results could be used not only 
in the specific cases discussed in the present paper. 

The plan of the paper is the following. 
In Section II we will review the basic features of blue 
spectra arising in quintessential inflationary models. In Section III 
we will set up the basic definitions and conventions concerning 
the evaluation of the SNR. In Section IV we will be mainly concerned 
with the analysis of the achievable sensitivities to some specific 
theoretical model.
Section V contains our concluding remarks.

\renewcommand{\theequation}{2.\arabic{equation}}
\setcounter{equation}{0}
\section{Blue and violet graviton spectra}

\subsection{Basic bounds}
Blue and violet logarithmic energy spectra of relic gravitons are 
phenomenologically allowed \cite{20}. At low frequencies the 
most constraining bound \cite{21} comes from the COBE observations 
\cite{22} of the first (thirty) multipole moments of the temperature 
fluctuations in the microwave sky which implies that 
$h^2_0\,\Omega_{{\rm GW}}(f_0,\eta_0)$ has to be smaller than 
$ 6.9 \times 10^{-11}$ for 
frequencies of the order of $H_0$. At intermediate frequencies
(i.e., $ f_{\rm p} \sim 10^{-8}$ Hz) the pulsar timing 
measurements \cite{23} imply that 
$\Omega_{\rm GW}(f_{\rm p}, \eta_0)$ should not exceed $10^{-8}$. 
In order to be compatible with the homogeneous and isotropic 
nucleosynthesis scenario \cite{24,25} we should  require that 
\begin{equation}
h^2_0\,\int \,\Omega_{\rm GW}(f, \eta_0)\;{\rm d} \ln{f}\,<\,0.2\,
\Omega_{\gamma}(\eta_0)\,h_0^2\,\simeq\,5\,\times\,10^{-6},
\label{ns}
\end{equation}
where $\Omega_{\gamma}(\eta_0)= 2.6 \times 10^{-5} ~h^{-2}_0$ 
is the fraction of critical energy density 
in the form of radiation at the present observation time. In
Eq. (\ref{ns}) the integral extends over all the modes present 
inside the horizon at the nucleosynthesis time. 
In the case of blue and violet logarithmic energy spectra the 
COBE and pulsar bounds are less relevant than the nucleosynthesis 
one and it is certainly allowed to have growing spectra 
without conflicting with any of the bounds.\footnote{Notice that 
the nucleosynthesis bound refers to the case where the underlying 
nucleosynthesis model is homogeneous and isotropic. The presence of 
magnetic fields and/or matter--antimatter fluctuations can 
slightly alter the picture \cite{26,27}.}

\subsection{Quintessential Spectra}

Recent measurements of the red-shift luminosity relation in type Ia 
supernovae \cite {28} suggest the presence of an effective cosmological term 
whose energy density can be as large as $0.8$ in critical units. 
Needless to say that this energy density is huge if compared with 
cosmological constant one would guess, for instance, from electroweak 
(spontaneous) symmetry breaking, i.e., 
$\rho_{\Lambda} \sim (250~{\rm GeV})^4$. In  order to cope with this problem 
various models have been proposed \cite{29} and some of them rely on 
the existence of some scalar field (the quintessence field) whose 
effective potential has no minimum \cite{30}. 
Therefore, according to this proposal the evolution of the quintessence 
field is dominated today by the 
potential providing then the wanted (time-dependent) cosmological term. 
In the past the evolution of the quintessence field is in general not dominated
by the potential. The crucial idea behind quintessential inflationary models 
is the identification of the inflaton $\phi$
 with the quintessence field \cite{11}. Therefore, 
the inflaton/quintessence  potential $V(\phi)$
 will lead to a slow-rolling phase 
of de Sitter type for $\phi <0$ and it will have no minimum for $\phi>0$. 
Hence, {\em after} the inflationary epoch (but {\em prior to} nucleosynthesis )
the Universe will be dominated by $\dot{\phi}^2$. This means, physically,
that the effective speed of sound of the sources driving the background 
geometry during the post-inflationary phase will be drastically different 
from the one of radiation (i.e., $c_{s} = 1/\sqrt{3}$ in natual units) 
and it will 
have a typical stiff form (i.e., $c_{s}=1$). The fact that in the 
post-inflationary phase the effective speed of sound equals the speed of light 
has important implications for the gravitational wave spectra as it was 
investigated in the past for a broad range of equations of state 
stiffer than radiation (i.e., $1/\sqrt{3} < c_{s} <1$) \cite{12}. 
The conclusion 
is that if an inflationary phase is followed by a phase whose effective 
equation of state is stiffer than radiation, then, the high frequency branch 
of the graviton spectra will grow in frequency. The tilt depends upon
the speed of sound and it is, in our notations, 
\begin{equation}
\alpha = \frac{6 c_{s}^2 - 2 }{3 c_{s}^2 + 1}.
\label{tilt}
\end{equation}
We can immediately see that for all the range of stiff equations of state 
(i.e., $1/\sqrt{3} < c_{s} <1$) we will have that $0 < \alpha<1$ . The case 
$\alpha=0$ corresponds to $c_{s}= 1/\sqrt{3}$. This simply means that if the 
inflationary phase is immediately followed by the ordinary radiation-dominated
phase the spectrum will be (as we know very well) flat. 
The case $c_s=1$ is the 
most interesting for the case of quintessential inflation. In this case the 
tilt is maximal (i.e., $\alpha=1 $). Moreover, a more precise   
calculation \cite{12,13} 
shows that the graviton spectrum is indeed logarithmically 
corrected as 
\begin{equation}
q(f,\eta_0) = \frac{f}{f_1} \ln^2{\biggl(\frac{f}{f_1}\biggr)}.
\label{qes}
\end{equation}
It is amusing to notice that this logarithmic correction occurs 
only in the case $c_{s}=1$ but not in the case of the other 
stiff (post-inflationary) background. 
The typical frequency $f_1(\eta_0)$ appearing in Eq. (\ref{qes}) 
is given, today, by
\begin{equation}
f_1(\eta_0) = 1132~N^{-\frac{1}{4}}_{s} 
\biggl(\frac{g_{{\rm dec}}}{g_{\rm th}}\biggr)^{\frac{1}{3}}~{\rm GHz}.
\label{fMax}
\end{equation}
Apart from the dependence upon the number of relativistic degrees of 
freedom (i.e., $g_{\rm dec} = 3.36$ and $g_{\rm th} = 106.75$ ) which 
is a trivial consequence of the red-sfhit, $f_{1}(\eta_0)$ does also 
depend upon $N_{s}$ which is the number of (minimally coupled) scalar 
degrees of freedom present during the inflationary phase.
The amplitude of the spectrum  depends upon $N_{s}$ as  
\begin{equation}
\overline{\Omega}(\eta_0) = \frac{1.64 \times 10^{-5}}{N_{s}}.
\end{equation}
The reason for the presence of $N_{s}$ is that 
all the minimally coupled scalar degrees of freedom 
present during the inflationary phase will be amplified sharing 
approximately the same spectrum of the two polarizations of the gravitons 
\cite{31}. The 
main physical difference is that the $N_{s}$ scalars are directly coupled to 
fermions and, therefore, they will decay and thermalize 
thanks to gauge interactions \cite{11}. If minimally coupled scalars would not 
be present (i.e., $N_{s}=0$)  the model would not be consistent since 
the Universe will be dominated by gravitons with (non-thermal) spectrum given 
by Eq. (\ref{qes}). The  energy density of the 
quanta associated with the minimally coupled 
scalars, amplified thanks to the background transition from the 
inflationary phase to the stiff phase, will decrease with the Universe 
expansion as $a^{-4}$ whereas the energy density of the background will 
decrease as $a^{-6}$. The moment at which the energy density 
of the background becomes sub-leading marks the beginning of the radiation 
dominated phase and it takes place at a (present) frequency of the order 
of the mHz \cite{13}. Notice that this frequency has been obtained 
by requiring the reheating mechanism to be only gravitational 
\cite{31}. This assumption migh be relaxed by considering 
different reheating mechanisms \cite{lin} (see also \cite{pb2}).
In order to satisfy the nucleosynthesis constraint in the framework 
of a quintessential model with gravitational reheating \cite{31}
we have to demand that 
\cite{11,13} 
\begin{equation}
\frac{3}{N_{s}} \biggl(\frac{g_{{\rm n}}}{g_{{\rm th}}}\biggr)^{1/3} < 0.07,
\label{req}
\end{equation}
where the factor of $3$ counts the two polarizations 
of the gravitons but also the quanta associated with the inflaton and 
$g_{{\rm n}} = 10.75$ is the number of spin degrees of freedom at $t_{{\rm n}}$.
For frequencies $f(\eta_0) >f_{1}(\eta_0)$ the spectra of the produced
gravitons are exponentially suppressed as $\exp{[- f/f_1]}$. This is 
a general feature of the spectra of massless particles produced thanks 
to the pumping action of the background geometry 
\cite{12}\footnote{Quintessential graviton spectra have, in general,
three branches: a soft branch 
(for $10^{-18}~{\rm  Hz}\;\laq\;f\;\laq\; 10^{-16}~{\rm Hz}$), 
a semi-hard branch 
(for $10^{-16}~{\rm Hz}\;\laq\;f\;\laq\;10^{-3}~{\rm Hz}$) 
and a hard branch which is the one mainly discussed 
in the present paper. The reason for this choice is obvious since 
the noise power spectra of the interferometric detectors are defined in a 
band which falls in the region of the hard branch of the theoretical 
spectrum.}

\renewcommand{\theequation}{3.\arabic{equation}}
\setcounter{equation}{0}
\section{Signal-to-noise ratio for monotonic blue spectra}

In order to detect a stochastic gravitational wave background in an optimal way
we have to correlate the outputs of two (or more) detectors \cite{17,18,19,40}. 
The signal received by a single detector can be thought as the 
sum of two components: the {\em signal} (given by 
the stochastic background itself) and the {\em noise} associated with each 
detector's measurement. The noise level associated with a single 
detector is, in general, larger than the expected theoretical signal.
This statement holds for most of the single (operating and/or foreseen)
gravitational waves detectors (with the possible exception of the LISA
space interferometer \cite{6}). Suppose now that instead of a single 
detectors we have a {\em couple} of detectors or, ideally, {\em a network} 
of detectors.
The signal registered at each detector will be 
\begin{equation}
s_{i} = h_{i}(t) + n_{i}(t),
\end{equation} 
where the index $i$ labels each different detector.
If the detectors are sufficiently far apart the ensamble average 
of the Fourier components of the noises is stochastically distributed which 
means that 
\begin{equation}
\langle n^{\ast}_{i}(f) n_{j}(f')\rangle = \frac{1}{2}\delta(f-f')
S^{(i)}_{n}(|f|), 
\end{equation}
where  $S_{n}(|f|)$ is the  one-sided noise power spectrum which is 
usually expressed in seconds. The very same quantity can be defined for
 the signal. By then assuming the noise levels 
to be statistically independent of the gravitational strain registered by the 
detectors we obtain Eq. (\ref{2}). 

Consider now the case of two correlated interferometers and define 
the following rescaled quantities:
\begin{itemize}
\item $\Sigma_n^{\,(i)}\,=\,S_n^{\,(i)}/S_0$ ($i\,=\,1,2$);
\item $\nu\,=\,f/f_0$;
\item $\Omega_{\rm GW}(f)\,=\,\Omega(f_{0})\,\omega (f)$. 
\end{itemize}
(in this Section we will not write the explicit dependence 
of the theoretical quantities upon $\eta_0$: they 
are meant to be considered at the present time).
Notice that $f_{0}$ is (approximately) the frequency where the noise 
power spectra are minimal and $\Omega_{\rm GW}(f_0)$ is the graviton 
(logarithmic) energy density at the frequency $f_0$.
 Therefore the signal-to-noise ratio can be expressed as:
\begin{equation}
{\rm SNR}^2 \,=\,\frac{3 H_0^2}{5 \sqrt{2}\,\pi^2}\;\sqrt{T}\;
\frac{\Omega_{\rm GW}(f_0)}{f_0^{5/2}\,S_0}\;J \;,
\end{equation}
where we defined the (dimension-less) integral 
\begin{equation}
J^2 \,=\,\int_0^{\infty}\,{\rm d} \nu\,
\frac{\gamma^2\,(f_0 \nu)\,\omega^2
 (f_0 \nu)}{\nu^6\,\Sigma_n^{\,(1)} (f_0 \nu)\, 
\Sigma_n^{\,(2)} (f_0 \nu)}\;.
\label{Jint}
\end{equation}
From this last expression we can deduce that the minimum detectable 
$h_{0}^2\,\Omega_{\rm GW}(f_0)$ is given by (1 yr = $\pi\,\times\,10^{7}$ s)
\begin{equation}
h_{0}^2\,\Omega_{\rm GW} (f_0) \simeq 4.0\,\times\,10^{32}\;
\frac{f_{0}^{5/2}S_0}{J}\;
\left(\,\frac{1\;{\rm yr}}{T}\,\right)^{1/2}\;{\rm SNR}^2
\end{equation} 
For example, by taking $f_0 = 100$ Hz and $S_0 = 10^{-44}$ s, we get 
\begin{equation}
h_{0}^2\,\Omega_{\rm GW} (100\,{\rm Hz})\,\simeq\,
\frac{4.0\,\times\,10^{-7}}{J}\;
\left(\,\frac{1\;{\rm yr}}{T}\,\right)^{1/2}\;{\rm SNR}^2\;.
\label{minsig}
\end{equation}
Therefore, the  estimate of the sensitivity 
of cross-correlation measurements between two 
detectors to a given 
spectrum $\Omega_{\rm GW} (f)$ reduces, in our case, 
to the calculation of the integral 
$J$ defined in Eq. (\ref{Jint}). Given a specific theoretical 
spectrum, $J$ can be  
 numerically determined  for the  wanted  pair of detectors.

\renewcommand{\theequation}{4.\arabic{equation}}
\setcounter{equation}{0}
\section{Achievable Sensitivities for quintessential spectra}

\begin{figure}
\centerline{\epsfxsize = 6.5 cm  \epsffile{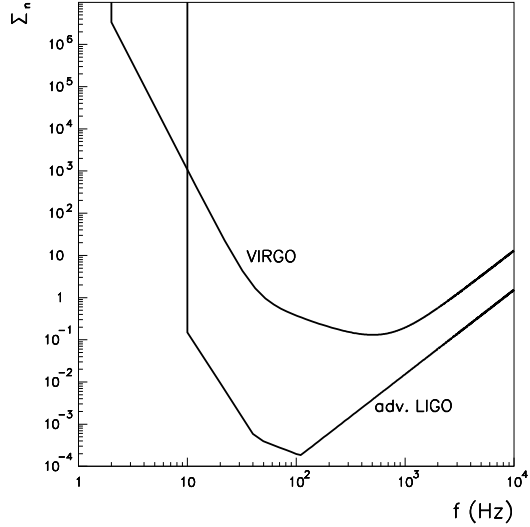}} 
\caption[a]{We report the rescaled noise power spectra of the LIGO and VIRGO 
detectors used for the calculation of the signal-to-noise ratio.}
\label{noise}
\end{figure}

Consider first the case of the two LIGO detectors (located at Hanford, WA and 
Livingston, LA) in their ``advanced'' versions. From the knowledge of the 
geographical locations and orientations of these detectors \cite{Allsit}, the 
overlap reduction function can be calculated \cite{19,40}, and the result is 
reported in Fig. 2 of ref. \cite{40}. 
As function of the frequency, $\gamma$ has 
its  first zero at 64 Hz and it falls rapidly at higher frequency. This 
behavior 
allows to restrict the integration domain in Eq. (\ref{Jint}) to the region 
$f \le 10$ kHz (i.e., $\nu \le 100$). We assumed identical construction of 
the two 
detectors (i.e., $S_n^{\,(1)} = S_n^{\,(2)}$). For the rescaled noise 
power spectrum 
of each detector we used the analytical fit of ref. \cite{41}, namely 
(see Fig. \ref{noise})
\begin{equation}
\Sigma_n (f)\,= \,
\left\{
\begin{array}{lc}
\infty & \qquad \qquad f < f_b \\ [8pt]
\dis h_a^2\,\left(\,\frac{f_a}{\Gamma}\,\right)^3\,\frac{1}{f^4} & 
\dis \qquad \qquad f_b \le f < \frac{f_a}{\Gamma} \\ [8pt]
\dis h_a^2 & \dis \qquad \qquad \frac{f_a}{\Gamma} \le f < \Gamma f_a \\ [8pt]
\dis \frac{h_a^2}{(\Gamma f_a)^3}\,f^2 & \qquad \qquad f \ge \Gamma f_a
\end{array}
\right.
\end{equation}
with
\bdis
h_a^2\,=\,1.96\,\times\,10^{-2} \quad \Gamma\,=\,1.6 \quad 
f_a\,=\,68\,{\rm Hz} \quad f_b\,=\,10\,{\rm Hz}.
\edis

In the case of a flat spectrum (i.e., $\alpha = 0$, $\omega(f) = 1$) we find 
$J\,\simeq\,6.1\,\times\,10^3$, which implies 
\begin{equation}
h^2_0\,\Omega_{{\rm GW}}(100\,{\rm Hz})\,\simeq\,6.5\,\times\,10^{-11}\;
\left(\,\frac{1\;{\rm yr}}{T}\,\right)^{1/2}\;{\rm SNR}^2
\label{flatligo}
\end{equation}
in close agreement with the estimate obtained in ref. \cite{40}.

The minimum detectable 
$h_{0}^2\,\Omega_{\rm GW}$ for quintessential gravitons can be obtained 
by recalling that 
\bdis
\omega (f)\,=\,\frac{\nu}{\ln^2 \nu_1}\;
\ln^2 \left(\,\frac{\nu}{\nu_1}\,\right).
\edis
For $f_0 = 100$ Hz, numerical integration gives:
\bdis
J\,\simeq\,\frac{10^3}{\ln^2 \nu_1}\;\left\{\,6.91\,+\,21.36\,\ln \nu_1\,+\,
26.52\,\ln^2 \nu_1\,+\,15.68\,\ln^3 \nu_1\,+\,3.78\,
\ln^4 \nu_1\,\right\}^{1/2}\;,
\edis
or, taking into account Eq. (\ref{fMax}), in terms of $N_s$:
\begin{equation}
J\,\simeq\,\frac{1.6\,\times\,10^7}{(88.0 - \ln N_s)^2}\;P_{{\rm L}} (N_s)
\end{equation}
with
\begin{eqnarray*}
P_{{\rm L}}^2 (N_s)\,&\simeq&\,1.07\,-\,4.62\,\times\,10^{-2}\,\ln N_s\,+\,
7.52\,\times\,10^{-4}\,\ln^2 N_s \\
& &\quad -\,5.44\,\times\,10^{-6}\,\ln^3 N_s\,+\,
1.48\,\times\,10^{-8}\,\ln^4 N_s\;.
\end{eqnarray*}
By inserting this expression in Eq. (\ref{minsig}), one has:
\begin{equation}
h_0^2\,\Omega_{\rm GW}(100\,{\rm Hz})\,\simeq\,2.5\,\times\,10^{-14}\;
\frac{(88.0 - \ln N_s)^2}{P_{{\rm L}} (N_s)}\;
\left(\,\frac{1\;{\rm yr}}{T}\,\right)^{1/2}\;{\rm SNR}^2
\label{poll}
\end{equation}
By assuming for $N_s$ the minimum value compatible with Eq. (\ref{req}) 
(i.e., $N_s = 21$), we obtain: 
\begin{equation}
h_0^2\,\Omega_{\rm GW}(100\,{\rm Hz})\,\simeq\,1.8\,\times\,10^{-10}\;
\left(\,\frac{1\;{\rm yr}}{T}\,\right)^{1/2}\;{\rm SNR}^2
\label{qligo}
\end{equation}
As we can see by comparing Eq. (\ref{flatligo}) with Eq. (\ref{qligo}) 
the minimum detectable $h_0^2\,\Omega_{\rm GW}(100\,{\rm Hz})$ is slightly 
larger for growing spectra. This a general result that is simply related to 
the structure of $J$. 
For the special value of $N_s$ considered this difference 
is roughly of a factor of 2. Another important point to stress 
is that for both 
the graviton spectra considered, as a consequence of the frequency behavior of 
$\gamma (f)$ and the presence of the weighing factor $\nu^{-6}$ 
in the integrand, the main contribution to the integral $J$ comes from the 
region $f < 100$ Hz. The cut-off introduced by the overlap reduction function 
is not so relevant: by assuming $\gamma (f) = 1$ over the whole integration 
domain (i.e., considering the correlation of one of the detector with itself), 
the sensitivity increases only by a factor 2.4 in the case of a flat spectrum, 
and 3.6 in the case of the quintessential one. This means that the only way 
to get a substantial rise in sensitivity lies in the improvement of the 
noise characteristics of the detectors in the low-frequency region.

\begin{figure}
\centerline{\epsfxsize = 6.5 cm  \epsffile{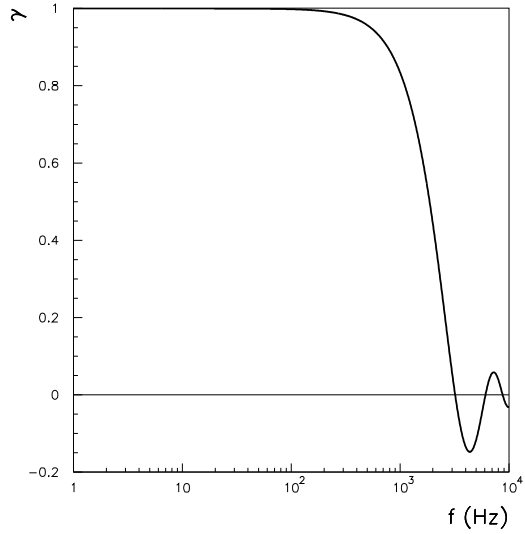}} 
\caption[a]{The overlap reduction function for the correlation of VIRGO with 
a coaligned 
interferometer whose central (corner) station is located at (43.2 N, 10.9 E), 
$d \simeq 58.0$ km (Italy).}
\label{fov}
\end{figure}

As a comparison we considered also the sensitivity that could be obtained at 
VIRGO in the (purely hypothetical) case in which the detector now under 
construction at Cascina, near Pisa (Italy), were correlated with a second 
interferometer located at about 50 km from the first and with the same 
orientation.\footnote{For illustrative purposes, 
we assumed, within our example,  50 km as the minimum distance sufficient to 
decorrelate local seismic and e.m. noises. This hypothesis might be proven 
to be correct and it is certainly justified in the spirit of this exercise.
However, at the moment, we do not have any indication either against or in 
favor of our assumption.} The overlap reduction function 
for this correlation has its 
first zero at a frequency $f \sim 3$ kHz (see Fig. \ref{fov}). 

Also in this case we assumed that the detectors are identical and for the 
common rescaled noise power spectrum we used the analytical parametrization 
given in ref. \cite{42} (see Fig. \ref{noise})
\begin{equation}
\Sigma_n (f)\,=\,
\left\{
\begin{array}{lc}
\infty & \qquad \qquad f < f_b \\ [8pt]
\dis \Sigma_1\,\biggl(\frac{f_{{\rm a }}}{f}\biggr)^5\,+\,
\dis \Sigma_2\,\biggl(\frac{f_{{\rm a}}}{f}\biggr)\,+\,
\dis \Sigma_3\,\biggl[ 1 + \biggl(\frac{f}{f_{\rm a}}\biggr)^2\biggr],& 
\qquad \qquad f \ge f_b
\end{array}
\right.
\end{equation}
where 
\bdis
f_a\,=\,500\,{\rm Hz}\;, \qquad f_b\,=\,2\,{\rm Hz}\;,\qquad
\begin{array}{c}
\Sigma_1\,=\,3.46\,\times\,10^{-6} \\
\Sigma_2\,=\,6.60\,\times\,10^{-2} \\ 
\Sigma_3\,=\,3.24\,\times\,10^{-2}
\end{array}
\edis
In the case of flat spectrum, limiting the numerical integration to 10 
kHz, 
we obtain $J\,\simeq\,5.5$ and, therefore, according to Eq. (\ref{minsig}) 
\begin{equation}
h_{0}^2\,\Omega_{\rm GW} (100~{\rm Hz})\,\simeq\,7.2\,\times\,10^{-8}\;
\biggl( \frac{1~{\rm yr}}{T}\biggr)^{1/2} 
{\rm SNR}^2\;.
\label{flatvirgo}
\end{equation}

In the case of quintessential inflation, for $f_0 = 100$ Hz, we have:
\bdis
J\,\simeq\,\frac1{\ln^2 \nu_1}\;\left\{\,5.79\,-\,0.30\,\ln \nu_1\,+\,
31.20\,\ln^2 \nu_1\,+\,6.11\,\ln^3 \nu_1\,+\,12.91\,
\ln^4 \nu_1\,\right\}^{1/2}\;,
\edis
or, in terms of $N_s$:
\begin{equation}
J\,\simeq\,\frac{1.6\,\times\,10^4}{(88.0 - \ln N_s)^2}\;P_{{\rm V}} (N_s)
\end{equation}
with
\begin{eqnarray*}
P_{{\rm V}}^2 (N_s)\,&\simeq&\,3.10\,-\,0.14\,\ln N_s\,+\,
2.37\,\times\,10^{-3}\,\ln^2 N_s \\
& &\quad -\,17.84\,\times\,10^{-6}\,\ln^3 N_s\,+\,
5.04\,\times\,10^{-8}\,\ln^4 N_s\;.
\end{eqnarray*}
Therefore, from Eq. (\ref{minsig}) one has:
\begin{equation}
h_0^2\,\Omega_{\rm GW}(100\,{\rm Hz})\,\simeq\,2.5\,\times\,10^{-11}\;
\frac{(88.0 - \ln N_s)^2}{P_{{\rm V}} (N_s)}\;
\left(\,\frac{1\;{\rm yr}}{T}\,\right)^{1/2}\;{\rm SNR}^2
\label{polv}
\end{equation}
that, for $N_s = 21$ gives: 
\begin{equation}
h_0^2\,\Omega_{\rm GW}(100\,{\rm Hz}) \,\simeq\,1.1\,\times\,10^{-7}\;
\left(\,\frac{1\;{\rm yr}}{T}\,\right)^{1/2}\;
{\rm SNR}^2
\label{qvirgo}
\end{equation}

\begin{figure}
\centerline{\epsfxsize = 6.5 cm  \epsffile{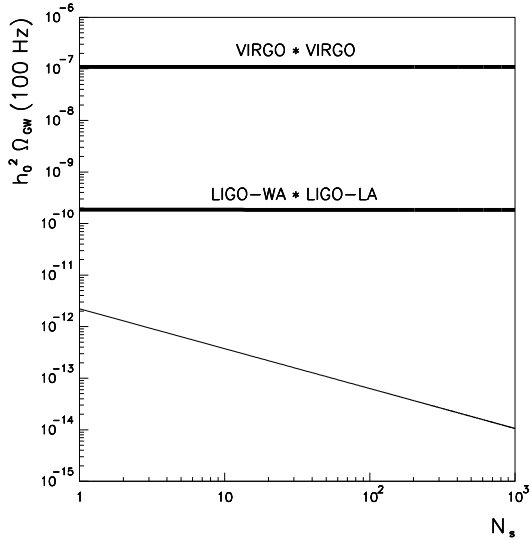}} 
\caption[a]{We report the theoretical
amplitude computed in Eq. (\ref{teorsign1}) (full thin line) and the 
associated sensitivities computed in Eqs. (\ref{poll}) and (\ref{polv}) 
for $T = 1$ yr and SNR = 1 (full thick lines). 
}
\label{f2}
\end{figure}

At a frequency of $100$ Hz the theoretical signal can be expressed as 
\begin{equation}
h^2_0\,\Omega_{\rm GW}(100\,{\rm Hz})\,=\,N_{s}^{-3/4}\, 
\times\,10^{-15}\;\biggl[2220.07\,-\, 50.46 \ln{N_{s}}\,+\,
0.28 \ln^2{N_{s}} \biggr]\;,
\label{teorsign1}
\end{equation}
as a function of $N_{s}$. In Fig. \ref{f2} this function (full thin 
line) is compared with the sensitivity of LIGO-WA*LIGO-LA and VIRGO*VIRGO 
(full thick lines) obtained from, respectively, Eqs. (\ref{poll}) and 
(\ref{polv}), assuming $T = 1$ yr and SNR = 1. We can clearly see that 
our signal is always below the achievable sensitivities. Notice that, if 
we assume purely gravitational reheating $N_{s}\;\gaq\;21$.

One could think that, thanks to the sharp growth of the spectrum, 
the signal could be strong enough around $10$ kHz, namely at the 
extreme border of the interferometers band. 
Indeed around  $f_0 = 10$ kHz, the theoretical signal 
is given by:
\begin{equation}
h^2_0\,\Omega_{\rm GW}(10\,{\rm kHz})\,=\,N_{s}^{-3/4}\, 
\times \,10^{-15}\;\biggl[1387.81\,-\,39.89 \ln{N_{s}}\,+\,
0.28 \ln^2{N_{s}} \biggr]\;,
\label{teorsign2}
\end{equation}
We see that the situation does not change qualitatively. In fact it is 
certainly true that around 
10 kHz the signal is  larger but the sensitivity is also 
smaller. In fact, repeating the calculation for $f_0 = 10$ kHz, in the case 
$N_s = 21$ we obtain: 
\begin{equation}
h_0^2\,\Omega_{{\rm GW}} (10\,{\rm kHz})\,\simeq\,
\left\{
\begin{array}{ll}
1.1\,\times\,10^{-8}\;\dis \left(\,\frac{1\;{\rm yr}}{T}\,\right)^{1/2}\; 
{\rm SNR}^2 & \qquad \qquad {\rm LIGO-WA}*{\rm LIGO-LA} \\
6.7\,\times\,10^{-6}\;\dis \left(\,\frac{1\;{\rm yr}}{T}\,\right)^{1/2}\; 
{\rm SNR}^2 & \qquad \qquad {\rm VIRGO}*{\rm VIRGO}
\end{array}
\right.
\label{q2vl}
\end{equation}
If we compare Eqs. (\ref{q2vl}) with Eqs. (\ref{qligo}) and (\ref{qvirgo}) 
we see that the minimum detectable signal gets larger the larger is the 
spectral frequency. 
Therefore, the mismatch appearent from Fig. \ref{f2} between the 
theoretical signal and the experimental sensitivity will remain practically
unchanged. 
 
\renewcommand{\theequation}{5.\arabic{equation}}
\setcounter{equation}{0}
\section{Concluding Remarks}

In this paper we precisely computed the sensitivity 
of pairs of interferometric 
detectors to blue and mildly violet spectra of relic gravitons.
Our investigation can be of general relevance for any model predicting non 
flat spectra of relic gravitons.  We analyzed the correlation 
of the two LIGO detectors in their ``advanced'' phase.  
On a more speculative ground we investigated the theoretical possibility 
of the correlation of VIRGO with an identical, coaligned, interferometer 
located very near to it. 

As a test for our techniques we first discussed 
the case of a flat spectrum which has been discussed in the past. We then 
applied our results to the case of quintessential inflationary models whose 
graviton spectra are, in general, characterized by three ``branches''.
A soft branch (in the far infra-red of the graviton spectrum around 
$10^{-18}$--$10^{-16}$ Hz), a semi-hard branch (between $10^{-16}$ Hz and 
$10^{-3}$ Hz) and a truly hard branch ranging, approximately, from 
$10^{-3}$ Hz to $100$ GHz.  Since the interferometers band 
is located, roughly, between few Hz and $10$ kHz, the relevant 
signal will come from the hard branch of the spectrum whose associated 
energy density appears in the signal-to-noise ratio 
with blue (or mildly violet) slope. In the hard branch
the energy density of quintessential 
gravitons is maximal for frequencies in the range of the GHz. In this 
region $h_0^2\,\Omega_{\rm GW}$ can be as large as $10^{-6}$. 
In spite of the fact quintessential spectra are growing in frequency the 
predicted signal is still too small and below the sensitivity achievable 
by the advanced LIGO detectors. 
The reason for the smallness of the signal in the region $f \sim 1$ kHz is 
twofold. On one hand we have to enforce the 
nucleosynthesis bound on the spectrum. 
On the other hand, because of the gravitational reheating mechanism 
adopted, the number of (minimally coupled) scalar degrees of freedom
needs to be large. It might be possible, in principle, that different 
reheating mechanisms could change the signal for frequencies comparable with 
the window of the interferometers.
Therefore, the analysis presented in this paper seems 
to suggest that new techniques (possibly based on electromagnetic 
detectors \cite{cav}) operating in the GHz region should be used 
in order to directly detect quintessential gravitons.

\section*{Acknowledgments}

We would like to thank Alex Vilenkin for very useful comments and 
conversations.

\end{document}